\documentclass[aps,pra,reprint, amsmath, amssymb,superscriptaddress, longbibliography]{revtex4-1}

\usepackage{bm}
\usepackage{bbm}
\usepackage[retainorgcmds]{IEEEtrantools}
\usepackage{graphicx}
\usepackage{mathrsfs}
\usepackage{amsmath}
\usepackage{amsfonts}
\usepackage{amssymb}
\usepackage{color}
\usepackage{times,txfonts}
\usepackage{nicefrac}
\usepackage{ragged2e} 
\usepackage{tikz}
\usepackage{physics} 
\usepackage{standalone} 
\usepackage{float}

\standaloneconfig{mode=buildnew}

\usepackage[colorlinks=true,linkcolor=blue,urlcolor=blue,citecolor=blue,pdfusetitle]{hyperref}

\usepackage{blindtext}

\definecolor{cadmiumgreen}{HTML}{097969}

\begin{document}

\title{Collisional thermometry for Gaussian systems}
\date{\today}

\author{Gabriel O. Alves}
\email{alves.go.co@gmail.com}
\affiliation{Max Planck Institute for the Physics of Complex Systems, 01187 Dresden, Germany}

\author{Marcelo A. F. Santos}
\email{augustofs.marcelo@gmail.com}
\affiliation{Instituto de Física Teórica, UNESP-Universidade Estadual Paulista, São Paulo 01140-070, São Paulo, Brazil}

\author{Gabriel T. Landi}
\email{gtlandi@gmail.com}
\affiliation{Department of Physics and Astronomy, University of Rochester, Rochester, New York 14627, USA}

\begin{abstract}

We investigate a quantum thermometry scheme based collision model with Gaussian systems. 
A key open question of these schemes concerns the scaling of the Quantum Fisher Information (QFI) with the number of ancillae. 
In qubit-based implementations this question is difficult to assess, due to the exponentially growing size of the Hilbert space. 
Here we focus on Gaussian collision models, which allow for the scaling of the QFI to be evaluated for arbitrarily large sizes. 
This numerical flexibility enables us to explore the thermometric properties of the model for a wide range of configurations. 
Despite the infinite Markov order of the stochastic process of the model, we provide a simple phenomenological analysis for the behavior of the QFI, estimating the asymptotic Fisher information density and how the transient effects of correlations for an increasing number of ancillae depend on the physical parameters of the model.

\end{abstract}

\maketitle{}


\section{Introduction}

The development of quantum technologies crucially depends on the precise control of quantum systems. 
The ability to extract information from them is necessary in order to perform practical tasks in several modern applications and quantum technologies \cite{jacobsQuantumMeasurementTheory2014, uchida_spintronic_2021, wuEnhancingQuantumEntanglement2023a, dekruijf2023measurement}
, with information theoretic approaches and continuous variable platforms being some of the prime examples \cite{braunsteinQuantumInformationContinuous2003, braunsteinQuantumInformationContinuous2005a, nokkalaGaussianStatesContinuousvariable2021b, spagnoloNonlinearBosonSampling2023}.
In that regard, quantum metrology emerges as the area of physics which employs and studies quantum systems as information probes \cite{giovannettiQuantumMetrology2006a, giovannettiAdvancesQuantumMetrology2011a}.

In a typical metrological setup, a quantum system is described by some parameter of interest. This parameter might be, for instance, the temperature of a condensate \cite{olfThermometryCoolingBose2015, mehboudiUsingPolaronsSubnK2019}, a phase \cite{hongQuantumEnhancedMultiplephase2021} or an external driving based on the electromagnetic field \cite{kaubrueggerOptimalVariationalMultiparameter2023}. 
In this sense, the relevant quantity in thermometry is the temperature of a bath coupled to a physical system. 
Equilibrium thermometry, where ancillae are let to fully thermalize with the system, is the standard template of such a scheme. In this design, the maximum achievable precision for unbiased estimators of the temperature is bounded bellow by the thermal Cramer-Rao thermal bound \cite{correa_individual_2015}.

A natural question which follows is whether it is be possible to surpass such limitation. In recent years, many works have tackled implementations of \emph{non-equilibrium} thermometry. The protocol known as collisional quantum thermometry came up as such an application, first appearing in \cite{Seah2019}, with further studies following thereafter \cite{shuSurpassingThermalCramerRao2020, oconnorStochasticCollisionalQuantum2021, boeyensUninformedBayesianQuantum2021, alves_bayesian_2022}. It was shown both that i) the non-equilibrium state of the probes resulted in enhanced thermal sensitivity, surpassing the thermal CRB, and also that ii) collective measurement of the ancillae could result in quantum advantages, surpassing the standard quantum limit (SQL) \cite{ hongQuantumEnhancedMultiplephase2021a}.


\begin{figure}
    \centering
    \includegraphics[width=\columnwidth]{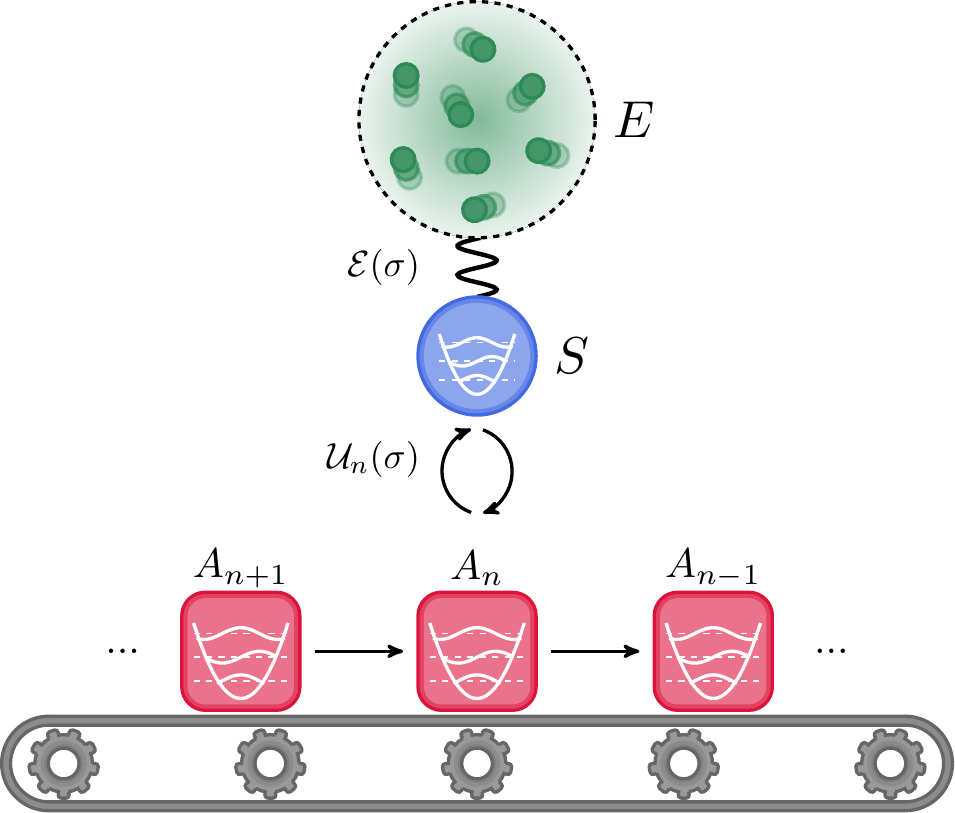}
    \caption{Depiction of the collisional thermometry scheme, based on bosonic systems. The Gaussian nature of the problem allows us to describe all the components in the model with a covariance matrices $\sigma$. The interaction between the intermediate system S and the environment E, and the system and the ancillae $A_n$ are given by the dissipative and unitary maps $\mathcal{E}(\sigma)$ and $\mathcal{U}_n$, respectively.}
    \label{fig:drawing_model}
\end{figure}

Meanwhile, an equally important venue of research is the non-asymptotic performance of metrological processes. Just beating the SQL by itself does not provide a complete picture -- or optimization -- of an experiment \cite{zwierzGeneralOptimalityHeisenberg2010c, escherQuantumMetrologyNoisy2011}. In that regard, much work has been done on how to efficiently perform parameter estimation on quantum systems in the limit of few measurements \cite{rubio_non-asymptotic_2018, rubioQuantumMetrologyPresence2019, yang_extractable_2023}. This is often closely tied with both Bayesian statistics \cite{rubio_global_2021} and with the study stochastic processes in more generality \cite{smiga_stochastic_2023}. In particular, a recent study \cite{radaelliFisherInformationCorrelated2023} has brought into light a few important aspects of correlations in stochastic processes of finite Markov order and how they impact estimation protocols.

Our objective in this manuscript is to further extend the collisional thermometry scheme to a model of Gaussian systems. However, rather than focusing on optimizing the protocol in typical settings, we concentrate instead on a few different aspects more related to how the correlations behave in this model and the transient aspects of the QFI, in the spirit of Ref.~\cite{radaelliFisherInformationCorrelated2023}. The reader interested in the former can encounter enlightening results in a recent work \cite{mirkhalafOperationalSignificanceNonclassicality2023}. With that in mind, we focus on a few scenarios where one introduces beam-splitter and two-mode squeezing interactions between bosonic ancillae and system, investigating a few figures of merit as these parameters are tuned.

One of our motivations for this choice of toy model lies in the fact that continuous variables, and oftentimes the subset of Gaussian systems and Gaussian operations, are the the natural description for several different physical platforms 
\cite{eisertGaussianQuantumChannels2005, monrasOptimalPhaseMeasurements2006, xueTwoModeSqueezedStates2007, weedbrookGaussianQuantumInformation2012, adessoContinuousVariableQuantum2014a, serafiniQuantumContinuousVariables2017a, lamiGaussianQuantumResource2018, lenziniIntegratedPhotonicPlatform2018, deoliveirajuniorUnravellingNonclassicalityRole2022}. 
On top of that, the extensive experimental expertise in the quantum physics community makes them an important object of study for quantum parameter estimation.
Their use and quantum advantages provided therein, have since long been of interest in many other subfields, such as quantum optics \cite{
cerfQuantumInformationContinuous2007a, zwierzUnifyingParameterEstimation2010a,
monrasMeasurementDampingTemperature2011b, 
ohOptimalGaussianMeasurements2019, michaelAugmentingSensingPerformance2021, kunduMachineLearningBasedParameterEstimation2022, 
kunduMachineLearningBasedParameterEstimation2022b, jonssonGaussianQuantumEstimation2022d},
and their use is naturally also very prevalent in quantum metrology as well
\cite{nicholsMultiparameterGaussianQuantum2018b, ohOptimalGaussianMeasurements2019c, liaoQuantumMetrologyMultimode2022a, naikooMultiparameterEstimationPerspective2023c}. 
Similarly, specific implementations of  thermometric protocols using continuous variables have recently gained traction 
\cite{monrasOptimalQuantumEstimation2007, monrasMeasurementDampingTemperature2011, correaEnhancementLowtemperatureThermometry2017b, ivanovQuantumThermometryTrapped2019, 
sekatskiOptimalNonequilibriumThermometry2022,
planellaBathInducedCorrelationsEnhance2022a, cenniThermometryGaussianQuantum2022b, jonssonGaussianQuantumEstimation2022, mirkhalafOperationalSignificanceNonclassicality2023, mihailescuThermometryStronglyCorrelated2023, parkQuantumLossSensing2023}, and familiar results have been extended to these cases. 
Finally, on the practical side of things, a Gaussian system with $N$ modes is fully described by a $2N \times 2N$ covariance matrix and a $2N$ dimensional vector of averages. This is in sharp contrast with qubit-based systems, where one obtains a exponential scaling with system size. Thus, besides the physical motivation,  Gaussian systems are possibly the simplest choice allowing feasible calculations even for large system sizes.
With this study we hope to supplement many of there aforementioned  investigations, further expanding the developments of quantum thermometry into the domain of bosonic systems, while at the same time shedding some light on the more detailed role of correlations in such processes. 

The paper is divided as follows: in Sec.~\ref{sec:collisional} we provide an overview of the collisional model using the Gaussian formalism. Sec.~\ref{sec:metrology} follows with a recapitulation of two important results and Gaussian metrology, where, in addition to that, we provide an efficient way of numerically computing the QFI in Sec.~\ref{sec:QFI_fast}. An analysis of the results is given in Sec.~\ref{sec:analysis}, with Sec.~\ref{sec:conclusion} concluding this manuscript.



\section{Collisional Model}\label{sec:collisional}


Collisional models consist in a scheme of repeated interactions between a system and identically prepared ancillae.
They initially appeared as a convenient tool to study open quantum system dynamics and, later on, have also been shown to be useful for many other applications, such as quantum thermodynamics and quantum information 
\cite{Rodrigues2019, Li2020, comarCorrelationsBreakingHomogenization2021a, liDissipationInducedInformationScrambling2022a, Camasca2021, ciccarelloQuantumCollisionModels2022a, filippovMultipartiteCorrelationsQuantum2022a}. See also Ref.~\cite{cusumanoQuantumCollisionModels2022a} for a review which provides a recent and pedagogical introduction to the topic.

In the collisional thermometry scheme, which is a generalization of standard probe-based thermometry \cite{correa_individual_2015}, a system S is let to interact with an environment E and a trail of independent and identically prepared (i.i.d.) ancillae $A_n$. As depicted in Fig~\ref{fig:drawing_model}, the system is placed in between a thermal bath at temperature $T$, the environment, and one of the ancillae $A_n$. 
The interactions happen as follows: the system S first undergoes a dissipative evolution through its interaction with E for a time $\tau_{SE}$, acquiring a temperature dependence on its state.  Afterward, the interaction is turned off and the system interacts solely with the Ancilla $A_n$ for time $\tau_{SA}$, in a unitary fashion. 
These alternating interactions are then repeated for the subsequent ancillae. This models a trail of ancillae that are used to probe the temperature of a thermal bath. 
We consider both the ancillae and the intermediate system to be bosonic modes, similarly as it has been done in \cite{Camasca2021} for the study of non-Markovianity. 
In other words, our model can be seen as a continuous variable implementation of the original proposal \cite{Seah2019}.


We restrict ourselves to the scenario where all of the ancillae are prepared in the same state $\rho_A^0$. If $\rho_S^0$ is the initial state of the in-between system, then the density matrix of the full system is given initially by
\begin{equation}\label{eq:initialstate}
    \rho^0 = \rho_S^0 \otimes \rho_A^0 \otimes \rho_A^0 \otimes \dots \otimes \rho_A^0.
\end{equation}
To simplify our analysis, we further constrain our model to Gaussian states. A Gaussian state of a bosonic system is completely determined by its first and second moments and thus has only a few degrees of freedom. These are the vector of means and the covariance matrix $\sigma$. The dynamics considered in this work are such that the mean is constant, and we thus set it to $0$ at all times without loss of generality. Meanwhile, the covariance matrix is defined as 
\begin{equation}
    \sigma_{ij} := \frac{1}{2}\langle {\hat{r}_i, \hat{r}_j} \rangle,
\end{equation}
where one defines $\hat{r} := (\hat{x}_S, \hat{p}_S, \hat{x}_{A_1}, \hat{p}_{A_1}, ... , \hat{x}_{A_N}, \hat{p}_{A_N})$ as the vector of canonical operators. Here the indices $S$ and $A_i$ indicate that the operator acts on the subspace of the system and the $i$-th ancilla, respectively. Meanwhile the position and momentum operators are defined in terms of the creation (annihilation) operator $a$ ($a^\dagger$) as: $\hat{x} := (a^\dagger + a)/\sqrt{2}$ and $\hat{p} := i(a^\dagger - a)/\sqrt{2}$.  The vector $\hat{r}$ also satisfies the commutation relation $[\hat{r}_i, \hat{r}_j] = i \Omega_{ij}$, where 
$\Omega = \bigoplus_i
\begin{pmatrix}
0 & 1 \\
-1 & 0
\end{pmatrix}
$
is the symplectic form \cite{adessoContinuousVariableQuantum2014a}.

For a product state, the covariance matrix is conveniently expressed as the direct sum of the covariance matrix of each component mode \cite{braskGaussianStatesOperations2022}. The covariance matrix of the initial state \eqref{eq:initialstate} is thus
\begin{equation}
    \sigma^{0} = \sigma_S^0 \oplus \sigma_A^0 \oplus \sigma_A^0 \oplus \dots \oplus \sigma_A^0.
\end{equation}

We begin by describing the unitary part of the evolution, which we denote as the system-ancilla (SA) interaction. The unitary interaction of the system between the $i$-th ancilla and the system is given by a combination of a beam-splitter (BS)
\begin{equation}\label{eq:BS_Hamiltonian}
    \hat{H}^i_{BS} = i g (a^{\dagger} b_i - a b_i^\dagger)
\end{equation}
and a two-mode squeezing (TMS) operation
\begin{equation}\label{eq:TMS_Hamiltonian}
    \hat{H}^i_{TMS} =  i h (a^{\dagger} b_i^{\dagger}- a b_i)  .
\end{equation}
This results in the total Hamiltonian $\hat{H}^i = \hat{H}^i_{BS} + \hat{H}^i_{TMS}$. Here, $a$ and $b_i$ denote the annihilation operator for the system mode and the $i$-th ancilla, respectively. 

At the level of the covariance matrix, the map associated with the interaction above is
\begin{equation}
    \sigma_{S A_i} \mapsto \mathcal{U}_{i}(\sigma_{S A_i}) = S_i \sigma_{S A_i} S_i^T,
\end{equation}
where $\sigma_{S A_i}$ denotes the joint CM of the system and the $i$-th ancilla, $S_i = e^{\Omega H_i \tau_{SA}}$ is the symplectic evolution operator and $H_i$ is the block matrix \cite{serafiniQuantumContinuousVariables2017a}
\begin{equation}\label{eq:ham_block_1}
    H
    =
    \begin{pmatrix} 
    0 & H_{SA} \\ 
    H_{SA} & 0
    \end{pmatrix},
\end{equation}
with
\begin{equation}\label{eq:int_hamiltonian}
    H_{SA} = 
    \begin{pmatrix} 
    g+h & 0 \\ 
    0 & g-h
    \end{pmatrix}.
\end{equation}


Meanwhile, the System-Environment (SE) interaction is an open system dynamics, described through a standard quantum master equation for the reduced density matrix of S,
\begin{equation}\label{eq:lindblad}
    \frac{d \rho_S}{dt} 
    = \gamma(\bar{n}+1)D_\rho[a] + \gamma \bar{n}D_\rho[a^\dagger],
\end{equation}
where $D_\rho[L] := L \rho L^\dagger  - \frac{1}{2}\{\rho, L^\dagger L\}$ is the dissipator and $\gamma$ is the coupling strength between the system and the bath. The action of this map alone drives the system to a thermal state $\sigma_{\mathrm th} := (\bar{n}+1/2) {\bm 1}$ in the long-time limit. This thermal evolution maps Gaussian states into Gaussian states and can be written  in terms of the covariance matrix as \cite{eisertGaussianQuantumChannels2005}
\begin{equation}\label{eq:thermalevolution}
\sigma_S \mapsto X\sigma_S X^T + Y,    
\end{equation}
where $\sigma_S$ is the reduced covariance matrix of the system,
\begin{equation}
    X 
    = 
    \exp\left(
    - \frac{\gamma \tau_{SE}}{2}
    \right) {\bm 1},
\end{equation}
and
\begin{equation}
    Y
    =
    \left(\bar{n} + 1/2\right)(1-e^{-\gamma \tau_{SE}}) {\bm 1}.
\end{equation}
The map in Eq.~\eqref{eq:thermalevolution} provides the evolution associated with the master equation~\eqref{eq:lindblad} in terms of the CM \cite{genoniConditionalUnconditionalGaussian2016}. 

We are interested in the composite application of the unitary and dissipative evolutions. This propels us to define a stroboscopic map \cite{molitorStroboscopicTwostrokeQuantum2020}:
\begin{equation}\label{eq:stroboscopicmap}
    \Phi(\sigma^{N-1})
    :=
    \sigma^N
    =
    \mathcal{U}_N (\mathcal{E}(\sigma^{N-1})),
\end{equation}
where $\sigma^n$ denotes the full covariance matrix after $N$ full steps of the collisional model.

We will perform our analysis based on the steady state $\sigma_S^*$ of the system associated with the map above. 
{
To do so, we first introduce
the Gaussian partial trace (over a covariance matrix).
This operation is defined as:
\begin{equation}
    \text{GTr}_{A} 
    \;
    \begin{matrix}
    \begin{pmatrix}
        \sigma_S && \sigma_{SA} \\
        \sigma_{SA}^T && \sigma_{A}
    \end{pmatrix}
    \end{matrix}
    :=
    \sigma_S,
\end{equation}
given the subsystems A and S and their correlation block $\sigma_{SA}$ \cite{serafiniQuantumContinuousVariables2017a}.
}
Now, note that the evolution of the in-between system in one time step is always given by
\begin{equation}
    \sigma_S \mapsto \text{GTr}_{A} \left[ S \left( \left(X \sigma_S X^T + Y \right) \oplus \sigma_A^0 \right) S^T \right].
\end{equation}
This is a consequence of the fact that the system always interacts with a fresh ancilla from the chain, in the initial state $\sigma_A^0$. 
The steady state $\sigma_S^*$ of the \emph{system} is obtained as the fixed point of this matrix evolution equation:
\begin{equation}\label{eq:steady_state}
    \sigma_S^* 
    = 
    \text{GTr}_{A} \left[ 
        S \left( \left(X \sigma_S^* X^T + Y \right) \oplus \sigma_A^0 \right) S^T 
    \right]
\end{equation}

The system is now allowed to evolve with $N$ steps of the collisional model. 
{
After this, the collective state of the system + ancilla is given by:
\begin{equation}\label{eq:stroboscopic_on_steady_state}
    \sigma^N = \mathcal{U}_N \circ \mathcal{E} \circ \dots \circ \mathcal{U}_1 \circ \mathcal{E} \; \sigma_0^*,
\end{equation}
with 
\begin{equation}
    \sigma_{0}^* := \sigma_S^* \oplus \sigma_A^0 \oplus \sigma_A^0 \oplus \dots \oplus \sigma_A^0.
\end{equation}
Here we have just replaced the (unimportant) initial state of the system with its steady-state $\sigma_S^*$ from Eq.~\eqref{eq:stroboscopic_on_steady_state}.
}
The final state of the $N$ ancillae, denoted $\Sigma^N$, is obtained by performing the partial trace over the degree of freedom of the system:
\begin{equation}\label{eq:state_n_ancillae}
    \Sigma^N = \text{GTr}_{S} \left[ \sigma^N \right].   
\end{equation}
This state contains information about the temperature of the bath. The correlations between ancillae are governed by the unitary contributions from Eqs.~\eqref{eq:BS_Hamiltonian} and \eqref{eq:TMS_Hamiltonian} and the choice of initial state. 

\section{Metrology for Gaussian systems}\label{sec:metrology}


\subsection{Quantum Parameter Estimation} 

\begin{figure}
    \centering
    \includegraphics[width=\columnwidth]{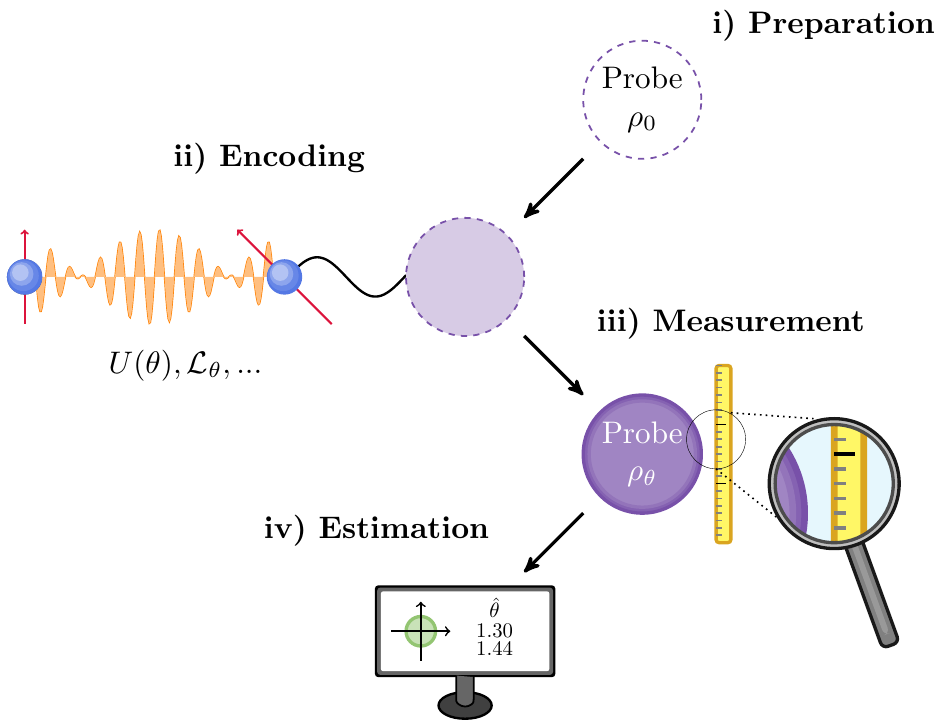}
    \caption{Diagram depicting the typical steps in quantum parameter estimation.}
    \label{fig:metrology_drawing}
\end{figure}

In this subsection we briefly review the general theory for quantum parameter estimation. The typical protocol is illustrated in Fig.~\ref{fig:metrology_drawing}.
To begin with, one prepares a probe in an arbitrary initial state $\rho_0$. 
The probe is nothing but a physical system used to encode a parameter of interest $\theta$.
Afterwards, it undergoes a dynamical process, embedding the parameter into the final state. 
In our particular case this corresponds to the open evolution in Eq.~\eqref{eq:lindblad}, which imprints the temperature dependence into the state of the system. 
Subsequently, the probe undergoes a measurement process where the experimental outcomes are obtained. Finally one employs these measurement results to estimate the parameter of interest. 

In the quantum setting there is, in principle, an infinite number of positive operator-valued measures (POVM) available. It is thus desirable to choose the best one.
When optimizing over all possible measurements, the maximum achievable precision associated with the mean-square error in this setting is given by the quantum Cram\'er-Rao bound (QCRB):
\begin{equation}\label{eq:QCRB}
\text{var }\hat{\theta} \geq \frac{1}{\mathcal{F(\theta)}},    
\end{equation}
which is valid for unbiased estimators \cite{Paris2009}. 
We call the quantity $\hat{\theta}$ an estimator. 
It is an arbitrary function used to used to estimate $\theta$ and depends solely on the experimental outcomes.  
The object $\mathcal{F}$ is the quantum Fisher information, defined as 
\begin{equation}\label{eq:QFI_standard}
    \mathcal{F}(\theta) := \tr{\rho_\theta L_\theta^2},
\end{equation}
and $L_\theta$ is the symmetric logarithmic derivative (SLD), an operator which is implicitly defined as the solution to the equation $\rho_\theta L_\theta + L_\theta \rho_\theta = 2 \partial_\theta \rho$. 

The QFI is a quantity which depends only on the final state $\rho_\theta$ and can, in principle, be readily calculated as long as $\rho_\theta$ is known. 
Eq.~\eqref{eq:QCRB} is quite useful in the sense that it makes no explicit reference to the particular choice of estimator or other post-processing aspects, so it helps to establish the ultimate achievable accuracy of a single-shot measurement with respect to the encoding state. 
Another important detail is that, although the QFI depends only on the encoded state at the end of the process, if the dynamics in \emph{step ii)} from the diagram is fixed, different choices of initial state yield different parametrizations. 
Thus, it is often a point of interest to also investigate how the protocol depends on the initial state. 
For instance, in Sec.~\ref{sec:squeezedthermometry} we will discuss the role of ancillae initialized in a squeezed state, as illustrated by the \emph{step i)} from the diagram. Meanwhile, in Sec.~\ref{sec:TMSthermometry} we fix the initial state $\rho_0$ and focus instead on how to optimize over the parameter dictating the dynamics, per \emph{step ii)}. 

As it can be seen from Eq.~\eqref{eq:QFI_standard}, the QFI and the SLD depend on the parameter itself. This often means that the accuracy of the estimation and the optimal basis will depend on the very value of what we are trying to estimate. 
This leads to two different perspectives or approaches, known as local and global metrology. The
latter normally makes use of the Bayesian framework, which was kick-started in \cite{rubio_global_2021} for thermometry and has been carefully investigated in other recent works \cite{Li2018, boeyensUninformedBayesianQuantum2021, alves_bayesian_2022,  morelliBayesianParameterEstimation2021a, jorgensenBayesianQuantumThermometry2022, mehboudiFundamentalLimitsBayesian2022}. 


\subsection{Gaussian metrology}

While the framework above is quite general, and Eq.~\eqref{eq:QCRB} allows us to calculate the QFI associated with any state and parameter, the calculations can quickly become burdensome in larger dimensional systems.
Considering that we are interested in Gaussian systems, it would be more useful to employ a formula which which can be written in terms of the CM and the momenta instead, properly making use of the Gaussian nature of the problem. 

A simple formula for the QFI of Gaussian systems has been derived in a previous work \cite{Monras2013}. There, it was shown that the QFI can be promptly calculated from the covariance matrix of a state and from the vector of first moments. In particular, whenever the first moments are zero, which was assumed in our case, the QFI associated with a Gaussian state $\sigma_\theta$ is given by the following expression:
\begin{equation}\label{eq:GaussianQFIApp}
\mathcal{F}(\theta) =  \frac{1}{2}\bra{\partial \sigma_\theta} \mathcal{D}(\sigma_\theta)^{-1} \ket{\partial \sigma_\theta}.
\end{equation}
Here we defined the operator:
\begin{equation}\label{eq:GaussianOperator1}
\mathcal{D}(C) := (C  \otimes C  + \Omega \otimes \Omega).
\end{equation}
A numerical maneuver which typically simplifies the computation is to rewrite the equation above as
\begin{equation}\label{eq:qfi_system}
\mathcal{F}(\theta) =  \frac{1}{2}\bra{\partial \sigma_\theta}  \ket{L_\theta},     
\end{equation}
where we define $\ket{L_\theta}$ as the solution to the linear system 
\begin{equation}\label{eq:QFIsystem}
\mathcal{D}(\sigma_\theta) \ket{L_\theta} = \ket{\partial \sigma_\theta}.
\end{equation}
This last step transforms the challenge of inverting the operator $\mathcal{D}(\sigma_\theta)$, into the problem solving a linear system for $\ket{\partial \sigma_\theta}$. In this scenario, numerical methods are readily available and very efficient with most standard libraries. This will therefore be faster than naively performing a matrix inversion directly. 

The expression in Eq.~\eqref{eq:GaussianQFIApp} is thoroughly equivalent to Eq.~\eqref{eq:QFI_standard}. Therefore, most of the computational work goes into the calculation of the inverse matrix above or the solution of the corresponding linear system. 
Nevertheless, it is clear that as one increases the number of ancillae, the dimensions in Eq.~\eqref{eq:GaussianQFIApp} increases \emph{quadratically} in $N$ due to the tensor product structure in the equation. 
In short, these alternative expressions avoid the hassle of dealing with infinite dimensional states by exploring the symplectic structure of the Gaussian systems.

\subsection{Efficient computation of the QFI for Gaussian systems}\label{sec:QFI_fast}

Here we introduce a trick which we will use to calculate Eq.~\eqref{eq:GaussianQFIApp}, showing that, by making use of Williamson's theorem \cite{pirandolaCorrelationMatricesTwomode2009a}, we arrive at an equivalent expression which is even less expensive computationally. 
Although the dimension of the CM increases linearly with the number of ancillae, calculating the QFI through Eqs.~\eqref{eq:GaussianQFIApp} or \eqref{eq:qfi_system} might still be a time-consuming task for arbitrary Gaussian states despite the massive speedup over equivalent expressions based on the full density matrix.
As mentioned in the previous paragraph, this is due to the fact that the dimensionality of the objects in Eq.~\eqref{eq:GaussianQFIApp} scales with $N^2$. 
Our strategy here is to first perform a symplectic diagonalization on the CM and all the other complicated operations in the original symplectic space of dimension $2N$. 
Once that is done, we can then re-express in Eq.~\eqref{eq:GaussianQFIApp} in terms of the symplectic eigenvalues. This will result in a very sparse matrix of fixed bandwidth, which virtually eliminates the drawback of working in this larger tensor product space associated with the complicated operator from Eq.~\eqref{eq:GaussianOperator1}.

We can begin by rewriting $\mathcal{D}(\sigma_\theta)$ as:
\begin{equation}\label{eq:GaussianOperator2}
\mathcal{D}(\sigma_\theta) = \frac{1}{2}(S \otimes S)\mathcal{D}(W)(S^T \otimes S^T).
\end{equation}
Here we made use of the fact that the symplectic matrix obeys $S \Omega S^T = \Omega$ and that the CM can be decomposed under a symplectic transformation $S$ as $\sigma = S W S^T$. 
The quantity $W$ is called the matrix of symplectic eigenvalues and it corresponds to the matrix obtained by the resulting diagonalization of the object $i \Omega \sigma$. That is:
\begin{equation}\label{eq:symplectic_eigs_matrix}
    W = \mathrm{diag}(\omega_1, ..., \omega_{2N}),
\end{equation}
given the eigenvalues $\{\omega_i\}$ of the matrix $i \Omega \sigma$. 
We can now rewrite Eq.~\eqref{eq:GaussianQFIApp} in terms of $W$ and the operator~\eqref{eq:GaussianOperator2}: 
\begin{equation}
\mathcal{F}(\theta) =  
\frac{1}{2}\bra{\partial \sigma_\theta} 
(S \otimes S)^{-T}
\mathcal{D}(W)^{-1} 
(S \otimes S)^{-1}
\ket{\partial \sigma_\theta},
\end{equation}
and then use the fact that $S^{-1} = - \Omega S^T \Omega$ to redefine a new vector $\ket{\overline{\partial \sigma_\theta}} \equiv (M \otimes M) \ket{\partial \sigma_\theta}$, where $M \equiv \Omega S^T \Omega$. This means that Eq.~\eqref{eq:GaussianQFIApp}  can be recast in the simpler form:
\begin{equation}\label{eq:QFI_efficient}
\mathcal{F}(\theta) =  \frac{1}{2}\bra{\overline{\partial \sigma_\theta}} \ket{\overline{L_\theta}}.     
\end{equation}
In analogy to Eq.~\eqref{eq:QFIsystem}, we define $\ket{\overline{L_\theta}}$ as the solution to the linear system 
\begin{equation}\label{eq:QFIsystem2}
\mathcal{D}(W) \ket{L_\theta} = \ket{\overline{\partial \sigma_\theta}}.
\end{equation}

\begin{figure}
    \centering
    \includegraphics[width=\columnwidth]{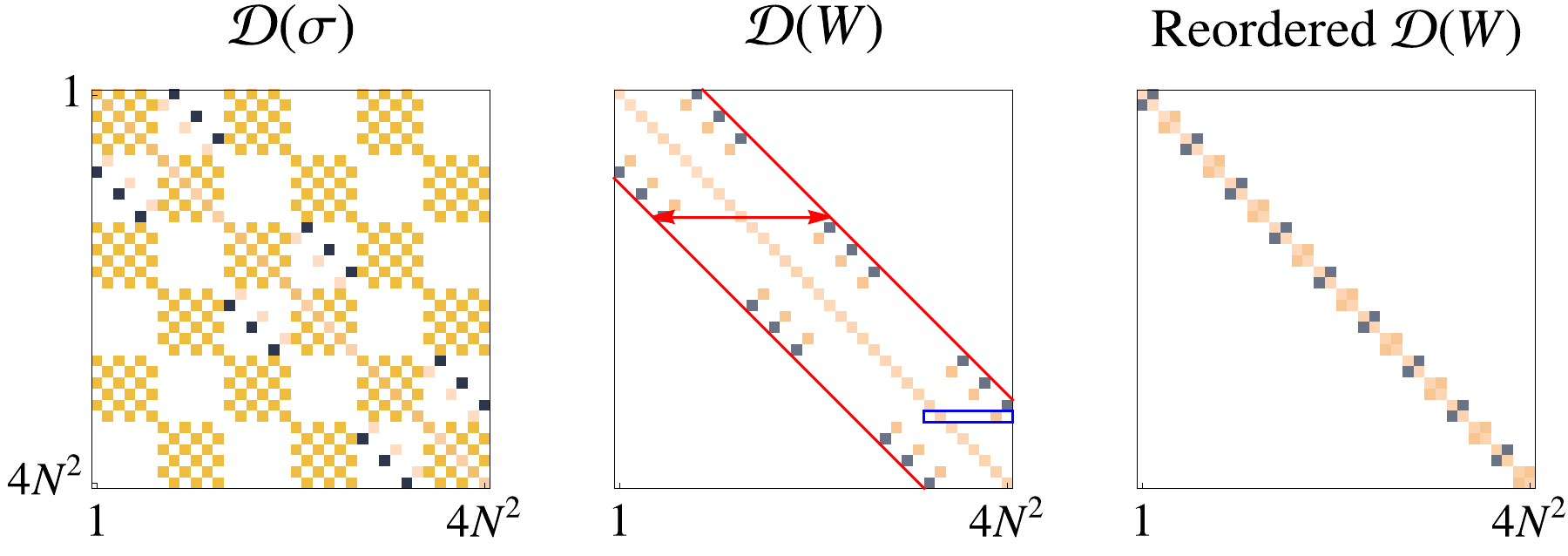}
    \caption{Illustration of the typical structure of the operator~\eqref{eq:GaussianOperator1} for the full CM $\sigma$ in this problem (left) and for the diagonal matrix of symplectic eigenvalues $W$ (right). Note how the operator on the right has a band structure. More importantly, this matrix is actually very sparse, since the number of elements in the band is always the same regardless of system size. The blue rectangle highlights the only two non-zero elements within the same row. Upon reordering of the matrix elements one can notice that this operator can actually be rewritten in terms of a diagonal composed of blocks of $2 \times 2$ matrices.}
    \label{fig:MatrixQFI}
\end{figure}

Finally, we explain the most important point about this trick, which makes our modification of Eq.~\eqref{eq:GaussianQFIApp} into Eq.~\eqref{eq:QFI_efficient}  provide a significant speedup for generic states.
The former expression requires us to invert a very dense matrix of dimensions $4N^2 \times 4N^2$. Since the typical algorithms for obtaining the inverse of matrices of size $m \times m$ are of complexity $O(m^3)$, this would translate into complexity $O(N^6)$ for a $N$-ancilla state. 
Meanwhile, the operator $\mathcal{D}(W)$ appearing in the modified expression is very sparse (since $W$ is diagonal) and has a band structure because of the term $\Omega \otimes \Omega$. This is illustrated on the middle panel of Fig.~\eqref{fig:MatrixQFI}. 
Thus, although we work with the same dimensionality in both equations, the latter is much easier to invert due to its simple structure
(c.f. Refs.~\cite{mahmoodFastBandedMatrix1991, asifBlockMatricesLblockbanded2005, ranInversionAlgorithmBanded2009, kilicInverseBandedMatrices2013}). 
However, this basic argument by itself is incomplete. 
Even more importantly, one last observation reveals that each row will always have a fixed number of non-zero elements regardless of the dimension of the matrix.
This can be directly verified by constructing the matrix $\mathcal{D}(W)$.
Upon close inspection we can see that by reordering this operator it is actually in block-diagonal form, composed of $2 \times 2$ blocks whose size does \emph{not} scale with $N$. This is shown in the rightmost panel of Fig.~\ref{fig:MatrixQFI}.
Since we have $2 N^2$ of these blocks, the number of operations for performing a LU decomposition and/or solving the linear system \eqref{eq:QFIsystem} is only of order $O(N^2)$. 

The whole point of this approach is, in other words, that instead of undergoing the numerical endeavor of inverting the generally dense matrix of dimensionality $4N^2$ appearing in Eq.~\eqref{eq:GaussianQFIApp}, we can focus  on obtaining the symplectic eigenvalues of the CM $\sigma_\theta$. 
Although this is not a trivial process and there is indeed an overhead, we are nevertheless performing all the heavy lifting in a space whose dimensionality is much smaller, since $\sigma_\theta$ has dimensions of $2N \times 2N$ and there is no tensor product structure at this point, in contrast to Eq.~\eqref{eq:GaussianOperator1}.
This approach should be valid for any parameter of interest $\theta$ and not just the temperature. 


\begin{figure*}[t]
    \centering
    \includegraphics[width=\textwidth]{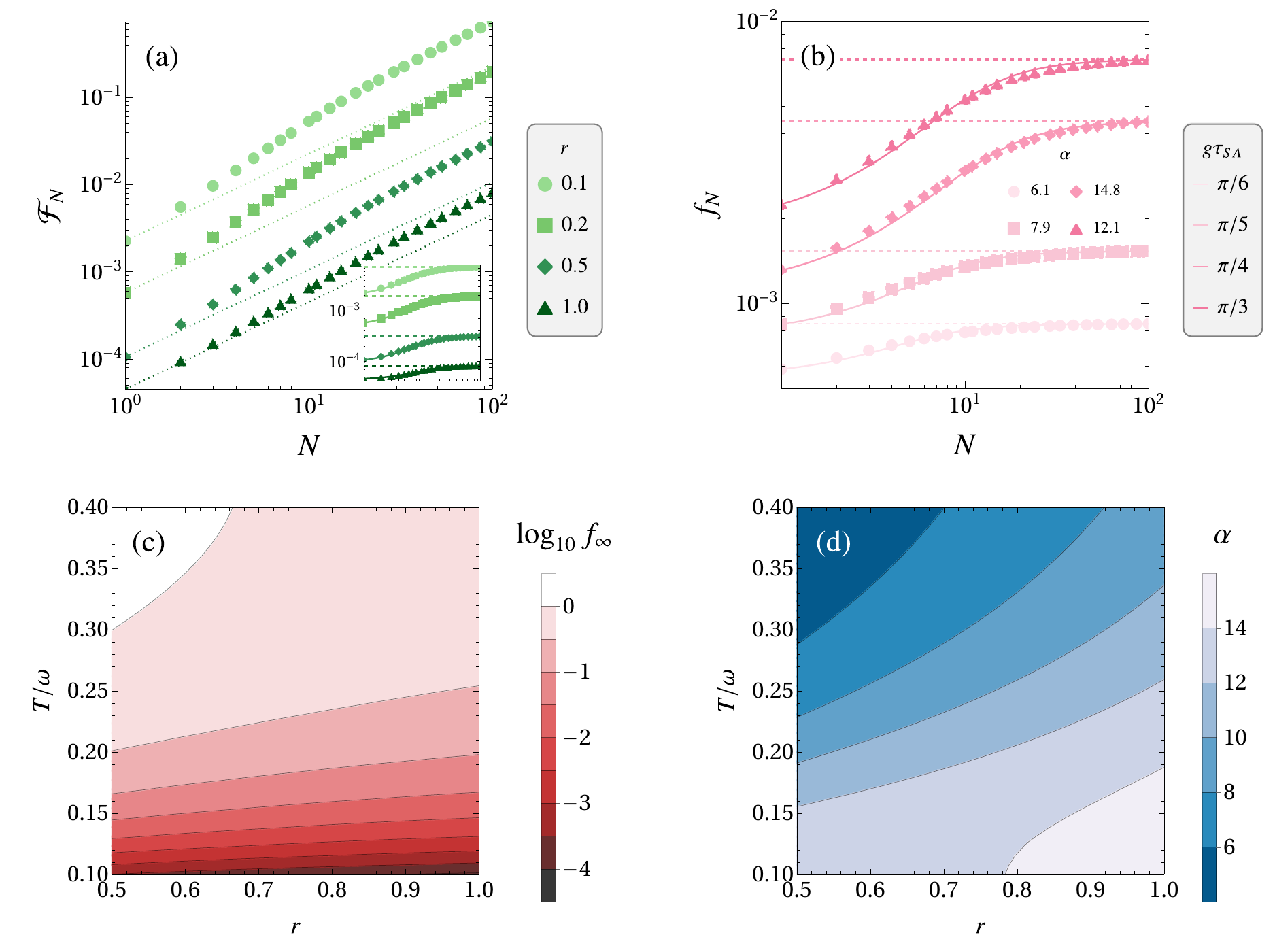}
    \caption{(a) QFI $\mathcal{F}_N$ for the state $\sigma_{A_1...A_N}$ as a function of $N$. Each curve corresponds to a different choice of squeezing parameter $r$. In the inset we plot the QFI density $f_N =\mathcal{F}_N/n$. (b) Fisher information density as function of $N$ for different BS couplings. The fitting parameter $\alpha$ for each different curve is shown in the plot. In panels (a) and (b) we fix $T/\Omega=0.1$. (c) Estimated QFI rate as function of the system parameters, for a fixed $g \tau_{SA} = \pi/3$. (d) Analogue plot where we show the fit parameter $\alpha$ as a function of the temperature and squeezing parameter. The fit is obtained by computing $f_N$ up to $n=200$. In all these figures we fix $\gamma \tau_{SE} = 1$.}
    \label{fig:QFISqueezingGrid}
\end{figure*}

\section{Analysis}\label{sec:analysis}

We are now equipped to discuss the numerical results of our simulations. We start by pointing that, differently from the qubit model where a partial-SWAP interaction alone provides super-linear scaling due to correlations between the ancillae \cite{Seah2019}, here we need some further ingredients for vacuum initialized probes.
More generally, the output of a beam-splitter field depends on the non-classicality of the input field~\cite{kimEntanglementBeamSplitter2002b, brunelliSingleTwomodeQuantumness2015a}.
In Sec.~\ref{sec:squeezedthermometry} we show how this can be achieved through single-mode squeezed states. 
Meanwhile, in Sec.~\ref{sec:TMSthermometry} we show how two-mode squeezing can achieve similar results. 
Note that in the latter case one should carefully discuss the stability and existence of the steady-state in Eq.~\eqref{eq:steady_state} \cite{Camasca2021}. 
We provide a brief discussion on this point in Appendix~\ref{app:steady_state}. In the simulations hereafter we shall always choose a set of parameters which result in a proper steady state.

\subsection{Squeezed ancillae thermometry}\label{sec:squeezedthermometry}

We start our analysis by considering the setup where the initial state of each ancilla is a squeezed vacuum state \cite{braskGaussianStatesOperations2022}:
\begin{equation}
    \sigma_{A}^0(r, \varphi) = 
    \frac{1}{2} \cosh (2r) \mathbbm{1}
    -
    \frac{1}{2} \sinh (2r)
    \begin{pmatrix}
        \cos \varphi & \sin \varphi \\
        \sin \varphi & - \cos \varphi
    \end{pmatrix}.
\end{equation}
For simplicity, we assume that $\varphi=0$ so that $\sigma_{A}^0(r, \varphi) = \sigma_{A}^0(r):= \text{diag }(e^{2r}, e^{-2r})$, since the phase rotations here can be disregarded~\cite{mirkhalafOperationalSignificanceNonclassicality2023}. 
To further simplify our analysis, {in this section} we also assume that no TMS is present, so we may take $h=0$ in Eq.~\eqref{eq:int_hamiltonian}, dispelling Eq.~\eqref{eq:TMS_Hamiltonian} from the dynamics. 
{We analyze the role of the TMS in Sec.~\ref{sec:TMSthermometry}}. 
Additionally, we keep the coupling strength $\gamma \tau_{SE}$ of the open evolution fixed for all simulations throughout the rest of this section.

The choices above mean that the tunable parameters in this first setting are (i) the squeezing parameter $r$ of the initial state of the ancillae, (ii) the strength $g \tau_{SA}$ of the BS interaction and (iii) the temperature $T$. The full-swap configuration with $g \tau_{SA} = \pi/2$ actually reduces to the partial thermalization process which has been investigated in Ref.~\cite{mirkhalafOperationalSignificanceNonclassicality2023}. 
The authors included a discussion about a (purposefully) more restrictive scenario, due to its great experimental relevance and tractability, where one is limited to Gaussian measurements only. They found that both single- and two-mode squeezing improve the accuracy of the thermometric scheme in the appropriate regimes.
In our case, what we do in this manuscript is to investigate how the QFI behaves as we vary these system parameters, giving a particular focus to how exactly it scales with $N$ for different system configurations. 

For that, we now describe the steps we follow in order to obtain the QFI for this model, summarizing the formalism from Secs.~\ref{sec:collisional} and ~\ref{sec:QFI_fast}. 
First, once all the parameters are chosen, we first obtain the steady-state of the system, given by Eq.~\eqref{eq:steady_state}. 
{After that we can apply the stroboscopic map~\eqref{eq:stroboscopicmap} to this steady-state, per Eq.~\eqref{eq:stroboscopic_on_steady_state}, obtaining the collective state of the sytem + ancillae.
Thus, after all the appropriate maps are applied, we can compute the collective CM of $N$-ancillae by simply tracing out the system, per Eq.~\eqref{eq:state_n_ancillae}. 
}
This will be the state of interest for the simulations. 
The reason is that we can now compute the symplectic eigenvalues of the corresponding CM in Eq.~\eqref{eq:state_n_ancillae}, leading us to the diagonal matrix $W$ in Eq.~\eqref{eq:symplectic_eigs_matrix}. 
Once that is done, it is then trivial to compute the QFI for the $N$-ancilla state by using Eq.~\eqref{eq:QFI_efficient}.

Now, we define the quantum Fisher information density (omitting the temperature dependence) as  $f_N := \mathcal{F}_N(T)/N$, where $\mathcal{F}_N(T)$ refers to the QFI of the collective state of $N$-ancillae~\eqref{eq:state_n_ancillae}, as described in the previous paragraph. 
Focusing on the QFI density will make some quantities and behaviors visually more apparent in upcoming figures and equations. 
Furthermore, our approach here was inspired by the recent work from Ref.~\cite{radaelliFisherInformationCorrelated2023}, where the authors were able to obtain a decomposition for the classical Fisher Information for a stochastic process of Markov order $\mathcal{M}$. 
They showed that for a process of Markov order $\mathcal{M}$, we can know everything about its metrological prowess for any longer sequence just by computing the FI up to $\mathcal{M} + 1$ outcomes.
Additionally, one of the points of investigation in their paper is that in such stochastic processes, one may observe a transient regime where the FI may be either super-additive or sub-additive, whereas the scaling eventually becomes linear once again for a large number of outcomes. 
This was verified both for certain classical Gaussian processes and for a Ising spin chain.

In analogy to the (classical) quantity of the same name in the aforementioned reference, we define here what we call the QFI rate:
\begin{equation}\label{eq:qfi_rate}
    f_\infty := \lim_{N \rightarrow \infty} f_N.
\end{equation}
This quantity is nothing but the value to which the QFI (asymptotically) converges to. 
For processes of finite Markov order this values is reached exactly after $\mathcal{M} + 1$ outcomes.
Unfortunately, in our case, there is no notion of a finite Markov order, so a decomposition as the one presented in Ref.~\cite{radaelliFisherInformationCorrelated2023} is not available, and Eq.~\eqref{eq:qfi_rate} is the only quantity which can be defined unambiguously. 
That is, there is no simple way to compute $\mathcal{F}_N$ in terms of the QFI for a smaller number of ancillae. On top of that, the authors in Ref.~\cite{radaelliFisherInformationCorrelated2023} also did not have the opportunity to investigate in detail how this dependence with $N$ in more general system with infinite Markov order, regardless whether classical or quantum. 
One of our objectives here is, among other things, to fill this gap with further numerical evidence and some supplemental analysis, finding out what is exactly the role of the correlations in this model at the level of the QFI density.

\begin{figure*}[ht!]
    \centering
    \includegraphics[width=\textwidth]{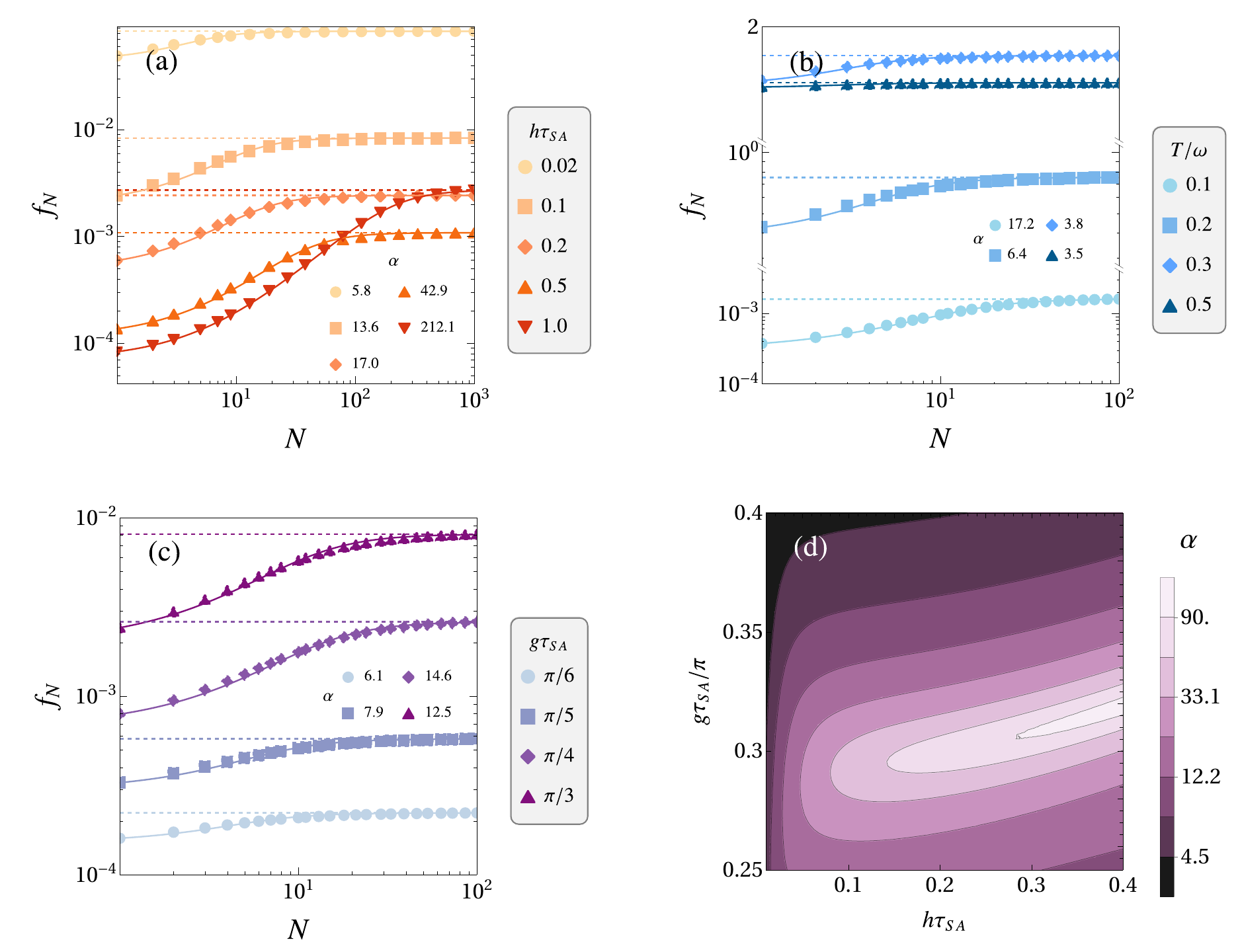}
    \caption{(a) QFI density as a function of $N$ for $g \tau_{SA} = \pi/3$, for different TMS couplings. We can observe that $h > 0$ introduces a transiently super-linear scaling for the QFI. (b) QFI density for different values of temperature. Note how the plot span across different scales in the vertical axis, showing how the fit provides a good approximation in all these cases. 
    (c) Analogous to panel (a). Here we show the QFI for different values of BS coupling, fixing $h \tau_{SA} = 0.1$.
    (d) Contour plot showing the fit parameter as function of the BS and TMS coupling. Note the log scale on the density map. Other details are chosen as in Fig.~\ref{fig:QFISqueezingGrid}.}
    \label{fig:QFIPlotGHGrid}
\end{figure*}

We can see a preliminary plot in the panel~(a) of Fig.~\ref{fig:QFIPlotGHGrid}, where we show the QFI and the QFI density~\eqref{eq:qfi_rate} (inset) as a function of $N$ for different values of the squeezing parameter in the initial state. 
This is the basic illustration which will allow us to visualize the transient behavior of the QFI. A first glance shows us two things: first, how the QFI quickly increases for the first few ancillae in a super-linear fashion, and second, how it eventually linearizes in an asymptotic manner. 
The inset make this point even clearer through the QFI density. 
For large $N$ the $f_N$ eventually saturates to an asymptotic value corresponding to the QFI rate from Eq.~\eqref{eq:qfi_rate}, which we highlight with the dashed lines in the figure.

Unfortunately, due to the complexity of these states, obtaining an analytical form for the QFI is out of reach even for a small number of ancillae. With that in mind, we opt for a more heuristic approach. In Fig.~\ref{fig:QFIPlotGHGrid}~(b) we once again plot the QFI density as a function of $N$, but this time we fix the squeezing parameter and vary the strength of the BS coupling. One can see that in either panels~(a) or~(b), $f_N$ follows a sigmoid-like behavior. For that reason, we propose a family single-parameter functions to fit this model, of the form:
\begin{equation}\label{eq:ansatz}
    f_{N + 1} = f_1 + (f_\infty - f_1)\frac{N}{\sqrt{\alpha^2 + N^2}}.
\end{equation}
Here $N$ is simply the number of ancillae, $f_\infty$ is the QFI rate and $f_1 = \mathcal{F}_1$ is the single shot QFI. The fittable parameter is $\alpha$. 
To approximate $f_\infty$ we simply compute $f_N$ for $N$ large.
This type of fit from Eq.~\eqref{eq:ansatz} reproduces almost exactly the behavior we will observe in our simulations. Note that when $N = 1$ we recover the single shot QFI, and that when $N \rightarrow \infty$ we recover the QFI rate. 
The factor $\frac{N}{\sqrt{\alpha^2 + N^2}}$ is simply an algebraic expression for a sigmoid function. Other common alternatives can also be used, such as a hyperbolic tangent of the form $\tanh (\alpha N)$. 
Those provide qualitatively similar results, but we have found that our current choice provided the best fit for the range of parameters considered here.

To illustrate, we show this function as a solid line in Fig.~\ref{fig:QFIPlotGHGrid}~(b) and the corresponding values of $\alpha$. Similarly, in the panels~(c) and~(d) we plot the two relevant quantities appearing in  Eq.~\eqref{eq:ansatz}. 
On panel~(c) we plot the QFI rate as a function of the temperature and the parameter $r$. 
As usual, we observe an exponentially vanishing value for the QFI as the temperature approaches zero. More interestingly, in panel~(d) we plot $\alpha$ as a function of $T$ and $r$. We obtained $\alpha$ by fitting Eq.~\eqref{eq:ansatz} for $200$ ancillae, which was enough in this setup to approach convergence. 
This figure shows the usefulness of our heuristic expression in Eq.~\eqref{eq:ansatz}: expressing the transient dynamics of the QFI ratio in terms of this single parameter $\alpha$ allows us to do so in a very condensed manner, with contour plots such as the one in panel~(d), which summarizes the transient behavior of the curves such as the ones in panels~(a) and ~(b).

Finally, we provide one last argument as to why we chose this approach. From the figures, the reader might notice that $\alpha$ provides a qualitative figure of merit describing \emph{how quickly} the QFI density reaches its asymptotic value, i.e., the memory effects for these states. 
This gives an operational meaning to the parameter $\alpha$: a larger value means that one should keep track of a larger number of ancillae in order to accurately describe $\mathcal{F}_N$ (or $f_N$).
Unfortunately, properly defining a notion of an \emph{approximated} Markov order is highly non-trivial \cite{fawziQuantumConditionalMutual2015, katoQuantumApproximateMarkov2019}, therefore we focus on a more naive characterization.
Nevertheless, to make our statement more precise, there is one last thing we can do. Assume that Eq.~\eqref{eq:ansatz} correctly describes the behavior of the QFI density. Now let us also suppose that we would like to obtain a QFI density of $(1 - \epsilon) f_\infty$, where $\epsilon$ roughly provides a notion of accuracy for an approximation of the QFI rate $f_\infty$. How many ancillae $N^*$ would be necessary for that? We simply solve the Eq.~\eqref{eq:ansatz} for $f_{N+1} = (1 - \epsilon)$, which will give us $N^*$. Through very simple algebra we obtain:
\begin{equation}
    N^*
    =
    (\alpha - 1)
    \frac{ f_1 + (1 - \epsilon) f_\infty }
    {\sqrt{f_\infty \epsilon [2 f_1 + (2 - \epsilon)f_\infty]}}.
\end{equation}
By assuming that $\epsilon \ll 1$, we get:
\begin{equation}
    N^* 
    \approx
    \frac{\alpha}{\sqrt{\epsilon}}
    \sqrt{\frac{f_\infty}{f_\infty - f_1}}.
\end{equation}
In particular, whenever $f_\infty \gg f_1$, we can see that $N^*$ scales with $N^*  \sim \alpha \epsilon^{-1/2}$. This means that, all other parameters fixed, the number of ancillae necessary to compute the QFI rate with precision $\epsilon$ is directly proportional to the fitting parameter $\alpha$. 
Also note how the limit $\alpha \rightarrow 0$ properly predict what we would expect, with Markov order zero. 
This corresponds to the case where no correlations are present, so Eq.~\eqref{eq:ansatz} is independent of $N$ and we have $f_{N+1} = f_1$. 
That is, the QFI $\mathcal{F}_N$ simply scales linearly.

Thus, our choice in Eq.~\eqref{eq:ansatz} gives a rough picture of how many ancillae one needs to consider when trying to compute the QFI for large $N$. Although the model has infinite Markov order strictly speaking, this type of picture provides an ad-hoc method of truncating the computation at a certain number $N^*$ of ancillae, after which the scaling can be assumed to be once again linear. 
This matches the intuition we can obtain from these pictures: even though the Markov order is infinite, we can still assume an approximately linear scaling after a certain number of ancillae is reached, approximately saturating $f_N$ to $f_\infty$. 
And what this figure show us is that how quickly this happens depends on the precise choice of parameters for the model. 
This is expressed, in our phenomenological analysis, through the parameter $\alpha$ in Eq.~\eqref{eq:ansatz}.

{ We also make a brief comment regarding the goodness of this fit. One can see that the fit might get slightly worse depending on the choice of parameters, but is unclear to us if or when this fit breaks down. 
We could not find a combination of parameters where this happens, so this is an open question left to more detailed investigations. 
More specifically, one could start by exploring different values of $\gamma \tau_{SE}$, which was kept fixed in our simulations.
Also, the fit might be a bit unstable numerically whenever $f_N$ is approximately constant, since certain numerical routines might be unable to distinguish whether $f_{\infty} \approx f_1$ or if $\alpha \rightarrow 0$.
}

\subsection{Two-mode squeezing thermometry}\label{sec:TMSthermometry}

TMS is a very well understood resource \cite{ riabininGeneratingTwomodeSqueezing2021b, cariolaroImplementationTwoModeGaussian2022b}
which has found a ubiquitous role in quantum information processing and quantum metrology 
 \cite{lauratEntanglementTwomodeGaussian2005b, 
anisimovQuantumMetrologyTwoMode2010a, jensenQuantumMemoryEntangled2011a,  liQuantumMetrologyTwomode2016a, cardosoSuperpositionTwomodeSqueezed2021a, souzaClassesGaussianStates2023a, liEquivalenceSqueezingEntanglement2023a, tritschler2023optical}.
For instance, Ref.~\cite{mirkhalafOperationalSignificanceNonclassicality2023} has showed that the TMS is able to improve the thermometric precision under certain situations. This motivates us to conclude our investigations by moving on to a second scenario, 
where we introduce TMS into the model. 
We follow all the steps in analogy with Sec.~\ref{sec:squeezedthermometry}, but this time turning on the TMS interaction appearing in Eqs.~\eqref{eq:ham_block_1} and~\eqref{eq:int_hamiltonian} from Sec.~\ref{sec:collisional}.

In this setup we initialize all the ancillae in the vacuum state $\sigma_0 = \frac{1}{2}\mathbbm{1}$ for simplicity, since the addition of TMS will now allow us to generate entanglement \cite{lauratEntanglementTwomodeGaussian2005c}. The tunable parameters become (i) the BS interaction and (ii) the temperature, as before, but this time we also have control over (iii) the TMS interaction through the coupling strength $h \tau_{SA}$. All other technicalities and assumptions are the same as in the previous section, unless clearly state otherwise. Our objective is now to repeat the same investigations but with a different set of parameters. The relevant plots are show in Fig.~\ref{fig:QFIPlotGHGrid}.

We start by fixing the BS coupling if Fig.~\ref{fig:QFIPlotGHGrid}~(a), where we plot the QFI density as a function of $N$ for different values of $h \tau_{SA}$.
We obtain a very similar behavior to the one observed in Fig.~\ref{fig:QFISqueezingGrid}, where we can see that the effects are even more pronounced. 
The solid line shows the numerical fit and the markers correspond to the numerical values. For the $h \tau_{SA} = 1$ (red downward markers) we obtain a large value for the fitting parameter, at $\alpha \approx 212$. Note how this is an order of magnitude larger than what we get for most of the other configurations and how this is very clear visually. We can see how the number of ancillae necessary to saturate to its asymptotic value is also much larger, around the order of $10^3$, while this happens around $10$ or $10^2$ for the other curves. 
Also note how we can easily perform these simulations even for $N$ or the order of $10^3$, showcasing the usefulness of the Gaussian formalism and the compact expression from Eq.~\eqref{eq:QFI_efficient}.
Obtaining similar plots for the qubit based model would be out of reach in generic scenarios.
{For instance, in Ref.~\cite{Seah2019} the authors investigate the scaling up to $12$ qubits.}

Finally, we do the same in Figs.~\ref{fig:QFIPlotGHGrid}~(b) and~(c), but varying the temperature and the strength, respectively. Note how the logistic-like behavior for these curves given in Eq.~\eqref{eq:ansatz} seem to the hold in all of these regimes. Meanwhile, the use of this fitting parameter once again allow us to summarize the information from Figs.~\ref{fig:QFIPlotGHGrid}~(a) and~(b) in a very condensed manner. In Fig.~\ref{fig:QFIPlotGHGrid}~(d) we plot $\alpha$ as a function of $g\tau_{SA}$ and $h\tau_{SA}$. 
Interestingly, this plot shows us that $\alpha \rightarrow 0$ as we approach the full-swap configuration $g \tau_{SA} \rightarrow \pi/2$, as it should be, since the correlations eventually vanish in this case. 
Additionally, also note that due to the presence of the TMS we lose the periodic behavior expected from the BS (or, analogously, the partial-swap interaction)~\cite{Seah2019}. 
Similarly, we can also see that there is local maximum for $\alpha$ in terms $g \tau_{SA}$ (within this regime) at each value of $h \tau_{SA}$, showcasing the very rich behavior of this model. 
What this panel is showing us is precisely a phenomenological picture of the model, where we can clearly see how different parameters strength of weaken these memory effects. By analyzing Fig.~\ref{fig:QFIPlotGHGrid}~(d) we can see, for instance, that as we increase the strength of the TMS or approach intermediate values for the BS interaction.
Such type of data basically provides a rule of thumb of how physically relevant the correlations present in the state~\eqref{eq:state_n_ancillae} are at the level of the quantum Fisher information. Or, similarly, how large should we take these collective ancillae states in order to accurately capture all the relevant information still with a good approximation. 
Taking $N^* \sim 10$ for $h \tau_{SA} = 0.02$ is already very representative. 
We can then assume that the scaling is linear after that.
Meanwhile, this would be disastrous for $h \tau_{SA} = 1$, where one would need to go to $N^* \sim 10^3$ as mentioned earlier. 


\section{Discussions and Conclusions}\label{sec:conclusion}

We have extended the original collisional thermometry proposal from Ref.~\cite{Seah2019} to an analogue based on Gaussian systems, which works as a  minimal model where we have included single-mode squeezed ancillae, beam-splitter interactions and two-mode squeezing. 
Inspired by recent advances in the study of correlated stochastic process and quantum metrology, we have used this model as a platform for focusing on the transient aspects and large-ancilla limit of the quantum Fisher information, rather than investigating features for optimal thermometry in the conventional sense.
We expect to have shed some light in some aspects which had not been  so much explored previously, specially in the context of quantum thermometry.
In most situation one cannot or should not expect an endless super-linear scaling, but rather, an intermediate behavior which eventually linearizes - as seen in Ref.~\cite{radaelliFisherInformationCorrelated2023}.  
It is then useful to figure out in more detail the rate at which a collective probe accumulates its metrological prowess due to memory effects. 
More concretely, our main objective here was to give a very clear picture of how exactly the QFI scales with the number of ancillae in this model. 
Our toy model was useful for that purpose, allowing us to operate with a wide range of parameters. 
Additionally, the alternative expression we presented here for the QFI of Gaussian states made these calculations feasible even for very large systems. 
However, it is not clear to us how universal this type of behavior is, and that lays the path for some further venues of investigation. 
One could for instance try to move on to more sophisticated bosonic (or fermionic) models used in condensed matter or quantum optics, checking whether an analogous behavior can still be observed there.

Finally, as stressed before, we have not touched upon other very important aspects in this type of problem, such as optimal preparation of probes and the effect of typically deleterious phenomena, such as decoherence, which can quickly compromise any naive protocol.
However, much progress has been made in this sense and one can try to extend ours  observations and any adjacent investigations into these other domains. 
See for example Ref.~\cite{baiFloquetEngineeringOvercome2023c} and the works mentioned therein. 
Similarly, it would be interesting to investigate whether the type of behavior and setup considered here has any type of implication for quantum many-body systems, since this type of collective state become the object of interest \cite{cieslinskiExploringManybodyInteractions2024}. 
Using this Gaussian state formalism may be particularly useful in this context, since one is naturally interested in large system sizes. This is strengthened by the fact that the QFI is actually an object of interest even beyond metrology. 
For instance, it may also serve as signature or diagnostic of other quantum effects and phenomena, such as  quantum scars~\cite{desaulesExtensiveMultipartiteEntanglement2022a} and quantum chaos~\cite{iniguezQuantumFisherInformation2023a}. 


\section*{Acknowledgments}

GOA thanks Pieter W. Claeys and Naim Elias Comar for the helpful discussions, and acknowledges the support by the Max Planck Computing and Data Facility and the financial support of the S\~ao Paulo Funding Agency FAPESP (Grant No.~2020/16050-0).
GTL acknowledges the financial support of the S\~ao Paulo Funding Agency FAPESP (Grants No.~2017/50304-7, 2017/07973-5 and 2018/12813-0), the Eichenwald foundation (Grant No.~0118 999 881 999 119 7253), and the Brazilian funding agency CNPq (Grant No. INCT-IQ 246569/2014-0). 

\appendix

\section{Stability of the steady state}\label{app:steady_state}

An analysis on the stability of fixed point of  map~\eqref{eq:stroboscopicmap} for the TMS and the BS unitaries can be found in Ref.~\cite{Camasca2021}. Here we extend this analysis to the case where both the TMS and BS are applied simultaneously. To do so, we first write Eq.~\eqref{eq:stroboscopicmap} explicitly using block matrices. The symplectic matrix $S$ that employs the unitary evolution of the system and one ancilla might be written

\begin{equation}
    S = \begin{pmatrix} 
    A & B \\ C & D 
    \end{pmatrix},
\end{equation}

\noindent where each submatrix is of dimension 2x2. Plugging this into Eq.~\eqref{eq:stroboscopicmap}, we obtain

\begin{equation}\label{eq:systemevol}
    \sigma_S^{n} = X(A \sigma_S^{n-1} A^T + B \sigma_A B^T)X^T + Y.
\end{equation}
Here, $\sigma_A$ is the state of the $N$-th ancilla. Note again that the ancillae are identically prepared.

Eq.~\eqref{eq:systemevol} can be solved for its steady state. First, for two any matrices $A$ and $B$, let us define the linear superoperator 
{$A\: \tilde{\otimes}\: B$ by $(A \otimes  B)C = A C B^T$, where $C$ is any other matrix.}
Then Eq.~\eqref{eq:systemevol} is, in this language, given by

\begin{align}
\begin{split}
    \sigma_S^n & = (X \: \tilde{\otimes} \: X)(A \: \tilde{\otimes} \: A) \sigma_S^{n-1} + (X \: \tilde{\otimes} \:X)(B \:\tilde{\otimes} \: B) \sigma_A + Y \\
\end{split}
\end{align}
We might recast this evolution equation in a vector form
\begin{equation}
    \sigma_S^{n} = \phi(\sigma_S^{n-1}) = G \sigma_S^{n-1} + H
\end{equation}
where $G$ and $H$ can be read off from the previous equation,
\begin{equation}
    G = (X \: \tilde{\otimes} \: X) (A \: \tilde{\otimes} \:  A), \quad H = (X \: \tilde{\otimes} \:  X)(B \: \tilde{\otimes} \:  B) \sigma_A + Y.
\end{equation}

Iterating this map we obtain an explicit expression for $\sigma^n_S$ in terms of the initial state $\sigma^0_S$,
\begin{equation}
    \sigma^n_{S} = G^n \sigma^0_S + \left( \sum_{k = 0}^n G^k \right) H.
\end{equation}
This evolution converges to the steady state $\sigma_{SS} = \phi(\sigma_{SS})$ for every initial condition iff the eigenvalues of $G$ all have an absolute value smaller than 1. 
For the case in hand, this simply implies that the eigenvalues of $AX$ have to be smaller than 1 in absolute value. 
The steady state might then be computed as
\begin{equation}
    \sigma_{SS} = (1 - G)^{-1}H.
\end{equation}
In particular, it does not depend on the choice of $\sigma^0_S$. 

\bibliography{library}

\end{document}